\documentclass[letter,11pt]{article}
\usepackage[top=1in,bottom=1in,left=1in,right=1in]{geometry}
\usepackage{slashed}
\usepackage{epsfig}
\usepackage{comment}
\usepackage{cancel}
\usepackage{bbm}
\usepackage{array}
\usepackage{bigints}
\usepackage{booktabs}
\usepackage{color}
\usepackage{dsfont}
\usepackage{float}
\usepackage{framed}
\usepackage{graphicx}
\usepackage{indentfirst}
\usepackage{mathrsfs}
\usepackage{multirow}
\usepackage{subdepth}
\usepackage{titlesec}
\usepackage[dotinlabels]{titletoc}
\usepackage{wrapfig}
\usepackage[all]{xy}
\usepackage[vcentermath]{youngtab}
\usepackage{relsize}
\usepackage{hyperref}
\numberwithin{equation}{section}

\usepackage[utf8]{inputenc}
\usepackage{slashed}
\usepackage{amsmath}
\usepackage{amsfonts}
\usepackage{amssymb}
\usepackage{cite}
\usepackage{mathtools}

\usepackage{cleveref}
\crefname{section}{§}{§§}
\Crefname{section}{§}{§§}

 \def\p{\partial}
 \def\bz{{\bar z}}
 \def\bw{{\bar w}}
 
\def\0{{(0)}}
\def\1{{(1)}}
\def\2{{(2)}}

\def\ci{{\mathscr I}}

\def\<{\langle }
\def\>{\rangle }
\def\bw{{\bar w}}
\newcommand{\bea}{\begin{eqnarray}}
\newcommand{\eea}{\end{eqnarray}}
\newcommand{\be}{\begin{equation}}
\newcommand{\ee}{\end{equation}}
\newcommand{\ba}{\begin{align}}
\newcommand{\ea}{\end{align}}

\renewcommand{\epsilon}{\varepsilon}

   \makeatletter
  \let\over=\@@over \let\overwithdelims=\@@overwithdelims
  \let\atop=\@@atop \let\atopwithdelims=\@@atopwithdelims
  \let\above=\@@above \let\abovewithdelims=\@@abovewithdelims
\renewcommand\section{\@startsection {section}{1}{\z@}%
                                   {-3.5ex \@plus -1ex \@minus -.2ex}
                                   {2.3ex \@plus.2ex}%
                                   {\normalfont\large\bfseries}}

\renewcommand\subsection{\@startsection{subsection}{2}{\z@}%
                                     {-3.25ex\@plus -1ex \@minus -.2ex}%
                                     {1.5ex \@plus .2ex}%
                                     {\normalfont\bfseries}}

\newcommand{\pd}[2]{\frac{\partial #1}{\partial #2}}

\newcommand{\beq}{\begin{equation}}
\newcommand{\eeq}{\end{equation}}
\newcommand{\beqa}{\begin{eqnarray}}
\newcommand{\eeqa}{\end{eqnarray}}
\newcommand{\beqar}{\begin{eqnarray*}}

\newcommand{\ve}{{\varepsilon}}

\def\[{\big[}
\def\]{\big]}

\def\ve{{\varepsilon}}

\def\g{{\gamma}}

\def\a{{\alpha}}
\def\b{{\beta}}

\def\bz{{\bar z}}

\def\be{{\bar \epsilon}}

\def\bw{{\bar w}}

\def\CA{{\mathcal A}}

\def\CC{{\mathcal C}}

\def\CG{{\mathcal G}}

\def\ci{{\mathcal I}}
\def\CJ{{\mathcal J}}

\def\CL{{\mathcal L}}

\def\CO{{\mathcal O}}

\def\CS{{\mathcal S}}

\newcommand{\avg}[1]{\langle\,   #1\,   \rangle}
\newcommand{\bra}[1]{\langle\,   #1\,    |}
\newcommand{\ket}[1]{ |\,   #1 \,  \rangle}
\newcommand{\braket}[2]{\langle\,   #1 \, | \, #2 \, \rangle}

\newcommand{\dt}{{\text d}}

\def\pmm{{(\pm)}}

\def\+{{(+)}}
\def\-{{(-)}}
\def\0{{(0)}}
\def\1{{(1)}}
\def\2{{(2)}}
\def\3{{(3)}}

\newcommand{\bd}[1]{\begin{fmffile}{#1}\begin{fmfgraph*}}
\newcommand{\ed}{\end{fmfgraph*}\end{fmffile}}

\begin{document}
\begin{titlepage}
\unitlength = 1mm
\ \\
\vskip 3cm
\begin{center}

{\LARGE{\textsc{A 2D Stress Tensor for 4D Gravity}}}

\vspace{0.8cm}
Daniel Kapec, Prahar Mitra, Ana-Maria Raclariu and Andrew Strominger

\vspace{1cm}

{\it  Center for the Fundamental Laws of Nature, Harvard University,\\
Cambridge, MA 02138, USA}

\vspace{0.8cm}

\begin{abstract}

We use the subleading soft-graviton theorem to construct an operator $T_{zz}$ whose insertion in the four-dimensional tree-level quantum gravity $\mathcal{S}$-matrix  obeys  the Virasoro-Ward identities of the energy momentum tensor of a two-dimensional conformal field theory (CFT$_2$).  The celestial sphere at Minkowskian null infinity plays the role of the Euclidean sphere of the 
CFT$_2$, with the Lorentz group acting as the unbroken $SL(2,\mathbb{C})$ subgroup.  

 \end{abstract}

\vspace{1.0cm}

\end{center}

\end{titlepage}

\pagestyle{empty}
\pagestyle{plain}

\def\vx{{\vec x}}
\def\p{\partial}
\def\po{$\cal P_O$}

\pagenumbering{arabic}
 

\tableofcontents

\section{Introduction}

Any quantum scattering amplitude of massless particles in four-dimensional (4D) asymptotically Minkowskian spacetime
can be rewritten as a correlation function on the celestial sphere at null infinity. Asymptotic one-particle states are represented as operator insertions on the sphere at the points where they  exit or enter the spacetime. The energy and other flavor or quantum numbers then label distinct operators.  The $SL(2,\mathbb{C})$ Lorentz invariance acts as the global conformal group on the celestial sphere and implies that these correlators lie in $SL(2,\mathbb{C})$ representations. 

In this paper we consider the $\CS$-matrix for 4D quantum gravity in asymptotically Minkowskian spacetime. We  construct an explicit soft-graviton mode, denoted $T_{zz}$,  and prove that its insertions in the tree-level $\CS$-matrix (with no other external soft insertions) obey all the Virasoro-Ward identities of a stress tensor insertion in a CFT$_2$ correlator on the sphere.  Our main tool is the subleading soft-graviton theorem \cite{Gross:1968in,Jackiw:1968zza,White:2011yy,Cachazo:2014fwa}. Our construction refines and extends results and conjectures of \cite{deBoer:2003vf,Banks:2003vp,Barnich:2009se,Barnich:2010eb,Barnich:2011ct,Kapec:2014opa}. It demonstrates that such quantum gravity scattering amplitudes are in Virasoro representations, as are CFT$_2$  correlators.  This extends from gauge theory to gravity earlier work  \cite{He:2014cra,He:2015zea} in which soft-photon and gluon insertions were shown to obey the Ward identities of a Kac-Moody algebra on the celestial sphere. 

The current work has several limitations. We do not consider massive particles, but do expect the extension to the massive case to be possible along the lines of \cite{Campiglia:2015qka,Kapec:2015ena,Campiglia:2015kxa}. Qualitatively important issues arise - including a possible central term - when there are multiple soft insertions that are not addressed here. At the one-loop level, corrections to the Ward identity are expected as a consequence of corrections to the soft theorem \cite{Bern:2014oka,He:2014bga,Bianchi:2014gla}. We have not analyzed their implications. Finally, although our results imply that certain quantum gravity scattering amplitudes are in Virasoro representations, there is no reason to expect that they are the same kinds of unitary representations appearing in conventional 2D CFTs.  We leave the nature of these representations to future work.

\section{Soft-Graviton Limits}

In this paper we consider tree-level scattering amplitudes of massless particles in four dimensions. The single particle states are labeled by $\ket{p,s}$, where $p$ and $s$ denote the four-momentum and helicity of the particle. The particles may carry charges or flavors but these indices are not relevant and are suppressed. The normalization of these states is given by
\begin{equation}
\begin{split}\label{statedef}
\braket{ p , s  }{ p' , s' } = (2\pi)^3 (2p^0) \delta_{s,s'} \delta^3 \big( \vec{p} - \vec{p}\,' \big) ~.
\end{split}
\end{equation}

The tree-level scattering amplitude involving $n$ massless states is denoted by\begin{equation}
\begin{split}\label{ampdef}
\CA_n = \bra{\text{out}} \CS \ket{\text{in} } ~, 
\end{split}
\end{equation}
where we use the shorthand $\ket{\text{in}}  = \ket{ p_1,s_1\,;\,\dots\,;\,p_m,s_m},~~ \bra{\text{out}}  = \bra{ p_{m+1},s_{m+1}\,;\,\dots\,;\,p_{n},s_{n}}$ and suppress the dependence of $\CA_n$ on the momenta $p_k$.   
We use a convention in which incoming states are described as CPT conjugate outgoing states with negative $p^0$ so that momentum conservation implies $\sum_{k=1}^np^\mu_k=0$.

Let $\CA^\pmm_{n+1}(q)$ be an amplitude involving a graviton of momentum $q^\mu$ and polarization $\ve^\pmm_{\mu\nu}(q)$ as well as $n$ other massless asymptotic states \begin{equation}
\begin{split}\label{ampwithgravitondef}
\CA^\pmm_{n+1}(q) =  \bra{\text{out}\,; \, q , \pm 2} \CS \ket{\text{in}} ~.
\end{split}
\end{equation}
 The soft $q^0 \to 0$ limit of this amplitude is governed by the leading \cite{Weinberg:1965nx} and sub-leading \cite{Gross:1968in,Jackiw:1968zza,White:2011yy,Cachazo:2014fwa} soft-graviton theorems\footnote{\label{footnote1}As shown in \cite{Cachazo:2014fwa,Broedel:2014fsa,Bern:2014vva}, tree-level graviton amplitudes are also constrained by a sub-subleading soft-graviton theorem.}
\begin{equation}
\begin{split}\label{softlimit}
\CA^\pmm_{n+1} (q) \to \left[ S^\pmm_0  + S^\pmm_1 + \CO  ( q  ) \right] \CA_n  ~, 
\end{split}
\end{equation}
where $\CA_n$ is the original amplitude without the soft-graviton \eqref{ampdef} and 
\begin{equation}
\begin{split}\label{softfactors}
S^\pmm_0 = \frac{\kappa}{2} \sum_{k=1}^n \frac{ p_k^\mu p_k^\nu  \ve^\pmm_{\mu\nu} (q)  }{ p_k \cdot q }  ~, \qquad S^\pmm_1 = - \frac{i\kappa}{2} \sum_{k=1}^n \frac{ \ve^{\pmm}_{\mu\nu}  (q) p_k^\mu  q_\lambda  }{ p_k \cdot q } \CJ_k^{\lambda\nu} ~, \qquad \kappa = \sqrt{32\pi G}~. 
\end{split}
\end{equation}
Here $\CJ_{k\mu\nu}$ is the angular momentum operator acting on the $k$th outgoing state. It is the sum of the orbital angular momentum operator $\CL_{k\mu\nu}$ and spin angular momentum $\CS_{k\mu\nu}$. Explicitly (see \cite{Weinberg:1995mt}),
\begin{equation}
\begin{split}\label{angmomop}
\CL_{k\mu\nu} &= - i \left[ p_{k\mu} \pd{}{p_k^\nu} - p_{k\nu} \pd{}{p_k^\mu } \right] ~, \\
\CS_{k\mu\nu} &=   - i s_k \left[    \ve^\+_\mu (p_k) \ve^\-_\nu (p_k)    -  \ve^\+_\nu (p_k)   \ve^\-_\mu (p_k)  \right] + s_k \ve^\+_\rho(p_k)   \CL_{k\mu\nu} \ve^{\-\rho} (p_k)   ~. 
\end{split}
\end{equation}
$\ve^\pmm_\mu(p)$ are polarization vectors that satisfy\footnote{Note that \eqref{polprop} is invariant under $\ve_\mu^\pmm (q) \to e^{i\theta_\pm(q) } \ve^\pmm_\mu(q)$, i.e. \eqref{polprop} only determines the polarizations up to an overall momentum dependent phase. These correspond to the little group transformations.}
\begin{equation}
\begin{split}\label{polprop}
 \ve^\pmm( p) \cdot p = 0 ~, \qquad  \ve^\pmm( p)\cdot \ve^\pmm ( p) = 0 ~, \qquad \ve^\pmm ( p) \cdot {\bar \ve}\,^\pmm ( p) = 1  ~. 
\end{split}
\end{equation}
Equation (\ref{angmomop}) continues to hold for particles of half-integer helicity provided that the little group phase of the wavefunction is chosen consistently. Gauge invariance of the leading and subleading soft limits implies momentum and angular momentum conservation respectively,
\begin{equation}
\begin{split}\label{conslaws}
\sum_{k=1}^n p_k^\mu \CA_n = \sum_{k=1}^n \CJ_{k\mu\nu} \CA_n = 0 ~. 
\end{split}
\end{equation}
To write out the soft factors explicitly, we parameterize the massless momenta and polarization vectors according to\footnote{In writing the explicit forms of the polarization vectors in \eqref{momexplicit}, we have specified our choice of little group phase. }
\begin{equation}
\begin{split}\label{momexplicit}
p_k^\mu &= \omega_k \left( 1 , \frac{z_k + \bz_k}{1 + z_k \bz_k} , \frac{ - i ( z_k - \bz_k )}{1 + z_k \bz_k} , \frac{1 - z_k \bz_k}{1 + z_k \bz_k} \right) ~, \quad k=1,\cdots,n  \\
\ve^\+_\mu( p_k ) &=  \frac{1}{\sqrt{2}} \left( - \bz_k , 1 , - i , - \bz_k \right)  ~,  \qquad \ve_\mu^\-(p_k )  =    \frac{1}{\sqrt{2}} \left( - z_k , 1 ,   i , - z_k \right)   ~,  \\
q^\mu &= \omega \left( 1  , \frac{z  + \bz}{1 + z  \bz}  , \frac{- i ( z  - \bz  )}{1 + z  \bz} , \frac{1 - z  \bz}{1 + z  \bz}   \right) ~, \\
\ve^\+_\mu(q) &=  \frac{1}{\sqrt{2}} \left( - \bz , 1 , - i , - \bz \right) ~, \qquad \ve_\mu^\-(q)  =    \frac{1}{\sqrt{2}} \left( - z , 1 ,   i , - z \right)   ~.
\end{split}
\end{equation}
The graviton polarization is $\ve^\pmm_{\mu\nu}(q) = \ve^\pmm_\mu(q) \ve^\pmm_\nu(q)$. In this parameterization, the soft factors \eqref{softfactors} are given by
\begin{equation}
\begin{split}\label{softfactorexplicit}
S_0^\+ &= - \frac{\kappa}{2\omega} \big(1+z\bz \big) \sum_{k=1}^n \frac{\omega_k ( \bz - \bz_k ) }{ ( z - z_k ) ( 1 + z_k \bz_k ) } ~, \\
S_0^\- &= - \frac{\kappa}{2\omega} \big(1+z\bz \big) \sum_{k=1}^n \frac{\omega_k ( z - z_k ) }{ ( \bz - \bz_k ) ( 1 + z_k \bz_k ) } ~, \\
S_1^\+ &= \frac{\kappa}{2} \sum_{k=1}^n  \frac{ ( \bz - \bz_k  )^2 }{ z - z_k  } \left[ \frac{ 2 {\bar h}_k  }{ \bz - \bz_k } -  \Gamma^{\bz_k}_{\bz_k \bz_k} {\bar h}_k  - \p_{\bz_k} + |s_k|   \Omega_{\bz_k} \right]  ~, \\
S_1^\- &= \frac{\kappa}{2} \sum_{k=1}^n  \frac{ ( z - z_k  )^2 }{ \bz - \bz_k  } \left[ \frac{ 2 h_k  }{ z - z_k } -  \Gamma^{z_k}_{z_kz_k} h_k  - \p_{z_k}  + |s_k|   \Omega_{z_k} \right]  ~. 
\end{split}
\end{equation}
Here $\Gamma^{z}_{zz}$ is the connection with respect to the unit round metric $\g_{z\bz} = 2  (1+z\bz )^{-2}$ on the sphere, $\Omega_z = \frac{1}{2} \Gamma^z_{zz} $ is the spin connection\footnote{The zweibein chosen here is $\big( e^+ , e^- \big)  = \sqrt{2\g_{z\bz}} \big( \dt z  , \dt \bz \big) $ for which $\Omega^\pm{}_\pm = \pm \frac{1}{2} \big(\Gamma^z_{zz} \dt z - \Gamma^{\bz}_{\ \bz\bz} \dt\bz \big)$. This choice is related to the little group phase chosen in \eqref{momexplicit}.}, and we have defined the operators\footnote{Single particle momentum eigenstates do not diagonalize the dilation operator $h_k+\bar h_k$. At tree-level, amplitudes are rational functions of the external momenta and we can formally define Mellin-transformed primary operators
$
\tilde{\mathcal{O}}(m,z,\bz)= \int_0^\infty d\omega \omega^{m-1}\mathcal{O}(\omega,z,\bz)
$
with conformal weights 
$
h=\frac12(s+m),\;\;\;\; \bar{h}=\frac12(-s+m)$.}
\begin{equation}
\begin{split}
h_k \equiv \frac{1}{2} \left( s_k - \omega_k \p_{\omega_k} \right) ~, \qquad {\bar h}_k \equiv \frac{1}{2} \left( - s_k - \omega_k \p_{\omega_k} \right) ~. 
\end{split}
\end{equation}
In this parameterization, equation \eqref{conslaws} takes the form
\begin{equation}
\begin{split}\label{conslawexplicit}
\left(\sum_{k=1}^n \omega_k\right) \CA_n =\left(\sum_{k=1}^n \omega_k\frac{z_k+\bz_k}{1+z_k\bz_k} \right)\CA_n = -i\left(\sum_{k=1}^n \omega_k\frac{ z_k-\bz_k }{1+z_k\bz_k}\right) \CA_n = \left(\sum_{k=1}^n \omega_k\frac{1- z_k \bz_k }{1+z_k\bz_k} \right)\CA_n &= 0 ~, \\ 
  - i \sum_{k=1}^n \left[ Y^{z_k} \big( \p_{z_k} - |s_k| \Omega_{z_k} \big)  + Y^{\bz_k}  \big( \p_{\bz_k}   - |s_k| \Omega_{{\bz_k}} \big)    +  D_{z_k} Y^{z_k} h_k   + D_{\bz_k} Y^{\bz_k} {\bar h}_k   \right]  \CA_n &= 0 ~, 
\end{split}
\end{equation}
where $Y^z(z) = a+bz+cz^2$ is a global conformal Killing vector and $D_z$ is the covariant derivative on the unit sphere.

\section{Mode Expansions Near $\ci^+$}
Four-dimensional asymptotically flat metrics
 \cite{Sachs:1962wk, Bondi:1962px,Barnich:2009se,Barnich:2010eb,Barnich:2011ct,Barnich:2011mi} 
admit an expansion near $\ci^+$of the form
\begin{equation}
\begin{split}
ds^2 &= - du^2 - 2 du dr + 2r^2\g_{z\bz} dz d\bz \\
&\qquad \qquad \qquad + \frac{2m_B}{r} du^2 + r C_{zz} dz^2 + r C_{\bz\bz} d\bz^2 + D^zC_{zz} du dz + D^{\bz}C_{\bz\bz} du d\bz + \cdots  ~. 
\end{split}
\end{equation}
In these coordinates $\ci^+$ is the null surface $(u,r=\infty,z,\bz)$. The retarded time $u$ parameterizes the null generators of $\ci^+$ and $(z,\bz)$ parameterize the conformal $S^2$. The boundaries of $\ci^+$ are located at $(u=\pm\infty,r=\infty,z,\bz)$ and are denoted $\ci^+_+$ and $\ci^+_-$ respectively.  The Bondi mass aspect $m_B$ and $C_{zz}$ depend only on $(u,z,\bz)$ and not on $r$. The news tensor is defined by
\begin{equation}
\begin{split}
N_{zz} \equiv \p_u C_{zz}~. 
\end{split}
\end{equation}
When expanding near flat spacetime, the Bondi coordinates are related to flat Cartesian coordinates by 
\begin{equation}
\begin{split}
x^0 = u + r ~, \qquad x^i = r {\hat x}^i(z,\bz) ~, \qquad {\hat x}^i(z,\bz) = \frac{1}{1+z\bz} \left( z + \bz , - i ( z - \bz ) ,1 - z \bz \right)~. 
\end{split}
\end{equation}
The space of asymptotically flat metrics in Bondi gauge with prescribed falloffs \cite{Sachs:1962wk, Bondi:1962px} admits an infinite-dimensional asymptotic symmetry group, the BMS group,  parameterized by vector fields of the form
\begin{align}
\xi  & = (1+\frac{u}{2r}) Y^{z}\p_z - \frac{u}{2r}  D^\bz D_z Y^{z} \p_{\bz} 
-\frac12 (u+r  )D_z  Y^{z} \p_r+{ u \over 2} D_z Y^{z} \p_u + c.c.\\ \notag
&\qquad +f\p_u  - \frac1r (D^zf\p_z + D^{\bz}f\p_{\bz}) +D^zD_zf\p_r + \cdots ~. 
\end{align}
Here $f(z,\bz)$ is a free function on the sphere associated to the supertranslation subgroup of the BMS group. The two-dimensional vector field $Y(z)$ is a conformal Killing vector (CKV) which realizes the action of the Lorentz group $SL(2,\mathbb{C})$ on the asymptotic sphere. 
For a more general CKV, obeying $\p_\bz Y^z=0$ except at isolated singularities, the Bondi gauge condition is preserved but the falloffs imposed on the metric are violated at the singularities. It was conjectured \cite{deBoer:2003vf,Banks:2003vp,Barnich:2009se,Barnich:2010eb,Barnich:2011ct} and proven in tree-level perturbation theory \cite{Kapec:2014opa} that such symmetries nevertheless play an important role.

The flat space outgoing graviton mode expansion is\footnote{Here, we take $g_{\mu\nu} = \eta_{\mu\nu} + \kappa h_{\mu\nu}$ which implies a canonical normalization for the graviton field, $\CL \sim \frac{1}{2} (\p h)^2$.}
\begin{equation}
\begin{split}
h_{\mu\nu}^{\text{out}} \big( x^0,\vec{x}\, \big) = \sum_{\a=\pm} \int \frac{d^3q}{(2\pi)^3} \frac{1}{2\omega_q} \left[ {\bar \ve}^{(\a)}_{\mu\nu} (q) a_\a^{\text{out}} ( q ) e^{ i q \cdot x } + \ve_{\mu\nu}^{(\a)} (q) a_\a^{\text{out}} (q)^\dagger e^{- i q \cdot x } \right] ~, 
\end{split}
\end{equation}
where $\omega_q = |\vec{q}\,|$ and
\begin{equation}
\begin{split}
\big[   a_\a^{\text{out}} ( p )   , a_{\b}^{\text{out}} (q )^\dagger \big] = \left( 2\pi \right)^3 \big( 2 p^0 \big)  \delta_{\a\b} \delta^3 \left( \vec{p} - \vec{q}\,\right) ~. 
\end{split}
\end{equation}
Outgoing gravitons with momentum $q$ and polarization $\alpha$ as in the amplitude \eqref{ampdef} correspond to final-state insertions of $a^{\text{out}}_\alpha ({\vec q})$.

In retarded Bondi coordinates
\begin{equation}
\begin{split}
C_{\bz\bz}(u,z,\bz) = \kappa \lim_{r\to\infty} \frac{1}{r} \p_{\bz} x^\mu \p_{\bz} x^\nu h_{\mu\nu}^{\text{out}} \big( u + r , r{ \hat x}(z,\bz) \big)  ~	. 
\end{split}
\end{equation}
This large $r$ limit can be computed using the stationary phase approximation \cite{He:2014laa,Kapec:2014opa} and one finds
\begin{equation}
\begin{split}\label{Czzexp}
C_{\bz\bz} (u,z,\bz) &=- \frac{i\kappa}{8\pi^2} {\hat \ve}_{\bz\bz} \int_0^\infty d\omega_q \left[ a^{\text{out}}_{-} \big( \omega_q {\hat x}\big) e^{- i \omega_q u } - a^{\text{out}}_{+} ( \omega_q {\hat x} \big)^\dagger e^{i \omega_q u} \right]  ~. 
\end{split}
\end{equation}
Here ${\hat x} \equiv {\hat x}(z,\bz)$ and 
\begin{equation}
\begin{split}
{\hat \ve}_{\bz\bz} = \frac{1}{r^2} \p_{\bz} x^\mu \p_{\bz} x^\nu \ve_{\mu\nu}^\+ (\omega_q {\hat x} )  = \frac{2}{(1+z\bz)^2}  ~. 
\end{split}
\end{equation}
Let us define
\begin{equation}
\begin{split}\label{Nzbzbomegadef}
N_{zz}^\omega  \equiv \int du e^{i\omega u} N_{zz}  ~, \qquad N_{\bz\bz}^\omega  \equiv \int du e^{i\omega u} N_{\bz\bz}   ~. 
\end{split}
\end{equation}
Then from \eqref{Czzexp}, we find for $\omega > 0$,
\begin{equation}
\begin{split}
N_{zz}^\omega &= - \frac{\kappa}{4\pi} {\hat \ve}_{zz} \omega a_{+}^{\text{out}} \big( \omega {\hat x} \big) ~, \qquad N_{zz}^{-\omega} = - \frac{\kappa}{4\pi} {\hat \ve}_{zz} \omega a_{-}^{\text{out}} \big( \omega {\hat x} \big)^\dagger ~, \\
N_{\bz\bz}^\omega &= - \frac{\kappa}{4\pi} {\hat \ve}_{\bz\bz} \omega a_{-}^{\text{out}} \big( \omega {\hat x} \big) ~, \qquad N_{\bz\bz}^{-\omega} = - \frac{\kappa}{4\pi} {\hat \ve}_{\bz\bz} \omega a_{+}^{\text{out}} \big( \omega {\hat x} \big)^\dagger ~. \\
\end{split}
\end{equation}
We now define the zero modes
\begin{equation}\begin{split}
N_{zz}^\0  &\equiv  \int du N_{zz} =\frac{1}{2} \lim_{\omega \to 0} \big( N_{zz}^\omega  + N_{zz}^{-\omega} \big) \\
&=  - \frac{\kappa}{8\pi} {\hat \ve}_{zz} \lim_{\omega \to 0} \big[ \omega a_{+}^{\text{out}}  \big( \omega {\hat x} \big) + \omega a_{-}^{\text{out}} \big( \omega {\hat x} \big)^\dagger   \big]  \end{split}\end{equation}
and 
\begin{equation}
\label{zeromodedef} \begin{split}
N_{\bz\bz}^\1 &\equiv  \int du u N_{\bz\bz} = - \frac{i}{2} \lim_{\omega \to 0} \p_\omega \big[ N_{\bz\bz}^\omega   - N_{\bz\bz}^{-\omega} \big]  \\
&=  \frac{i\kappa}{8\pi}  {\hat \ve}_{\bz\bz}  \lim_{\omega \to 0} \left( 1 + \omega \p_\omega \right)   \big[   a_{-}^{\text{out}} \big( \omega {\hat x} \big)  -    a_{+}^{\text{out}} \big( \omega {\hat x} \big)^\dagger  \big]  ~, \end{split}
\end{equation}
along with similar definitions for $N_{\bz\bz}^\0$ and $N_{zz}^\1$. We note that $N^\1_{\bz\bz}$ involves one less factor of $\omega$ than $N^\0_{zz}$, but has the Weinberg pole projected out by the factor of $1 + \omega \p_\omega $. Hence it has nonzero finite  scattering amplitudes.

The insertion of the zero mode \eqref{zeromodedef} is then given by \eqref{softlimit} and \eqref{softfactorexplicit} with
\begin{align}
\label{zeromodeins2}
 \bra{\text{out}} N_{\bz\bz}^\1 \CS \ket{\text{in}}  &= \frac{4 G i }{(1+z\bz)^2}  \sum_{k=1}^n  \frac{ (z - z_k  )^2 }{ \bz - \bz_k  } \left[ \frac{2 h_k }{z - z_k} -   \Gamma^{z_k}_{z_kz_k} h_k  - \p_{z_k}     +  |s_k| \Omega_{z_k}  \right]  \bra{\text{out}}    \CS \ket{\text{in}} ~. 
\end{align}

\section{A 2D Stress Tensor}

Massless scattering amplitudes $\CA_n$ of any four-dimensional theory may always be recast as two-dimensional correlation functions of local operators on the asymptotic $S^2$ at null infinity \cite{He:2015zea},
\begin{equation}
\begin{split}
\CA_n = \avg{ \CO_1(\omega_1,z_1,\bz_1) \cdots \CO_n(\omega_n,z_n,\bz_n) } ~. 
\end{split}
\end{equation}
The operator $\CO_k$ creates a massless single-particle state with momentum and polarization given by (\ref{momexplicit}).
The particle intersects the asymptotic $S^2$ at the point $(z_k,\bz_k)$\footnote{The same is not true for scattering amplitudes involving massive particles since a massive four-momentum does not localize to a point on $\ci$. However following \cite{Kapec:2015ena,Campiglia:2015qka,Campiglia:2015kxa} we expect the analysis of this paper to have a suitable generalization to the massive case, as the subleading soft theorem \cite{Gross:1968in,Jackiw:1968zza,White:2011yy,Cachazo:2014fwa} remains valid for massive particles.}. The four-dimensional Lorentz group $SL(2,\mathbb{C})$ acts as the global conformal group on the asymptotic $S^2$ according to\footnote{This also acts on the energy as
\begin{equation} \notag
\begin{split}
{\tilde \omega} \to  {\tilde \omega}  |cz+d|^2  ~, \qquad {\tilde \omega} = \frac{\omega}{1+z\bz} ~. 
\end{split}
\end{equation}
}
\begin{equation}
\begin{split}
z \to z' = \frac{a z + b }{ c z + d } ~, \qquad a d - b c = 1 ~. 
\end{split}
\end{equation}
This implies that all Minkowskian QFT$_4$ amplitudes are in representations of the same global conformal group as Euclidean CFT$_2$ correlators. In this section we will see that (hard) quantum gravity amplitudes are in representations of the full CFT$_2$ Virasoro group. Indeed it has already been shown that the leading soft-photon and graviton theorems are the Ward identities of abelian Kac-Moody current algebras acting on the asymptotic $S^2$ \cite{as,as1,He:2014laa,He:2014cra}. A similar Kac-Moody structure for non-abelian gauge theory scattering amplitudes was studied in \cite{Nair:1988bq}. The leading soft-gluon theorem in a non-abelian gauge theory with gauge group $\CG$ was shown in \cite{He:2015zea} to be equivalent to the Ward identity of a $\CG$ Kac-Moody current algebra. In all of these cases, holomorphic Kac-Moody current insertions were related to positive helicity soft insertions. For instance, the soft-photon Kac-Moody current is 
\begin{equation}
\begin{split}
J_z &=-\frac{8\pi}{e^2} F_{uz}^{(0)}  =   \frac{1}{e}  {\hat \ve}_{z} \lim_{\omega \to 0} \big[ \omega a_{+}^{\text{out}}  \big( \omega {\hat x} \big) + \omega a_{-}^{\text{out}} \big( \omega {\hat x} \big)^\dagger   \big]  ~,
\end{split}
\end{equation}
where $F_{uz}^{(0)}$ is the zero mode of the photon field strength, $ {\hat \ve}_{z}=\sqrt{ {\hat \ve}_{zz}}$, and $a_{+}^{\text{out}}  \big( \omega {\hat x} \big)$ creates outgoing positive helicity photons. Insertions of this current take the form
\begin{equation}
\begin{split}
\avg{ J_z \CO_1 \cdots \CO_n } &=   \sum_k \frac{Q_k  }{ z - z_k }  
\avg{ \CO_1 \cdots \CO_n  } ~, 
\end{split}
\end{equation}
where $eQ_k$ is the electric charge of the operator $\mathcal{O}_k$ and we have dropped the dependence of the operators on $(\omega_k,z_k,\bz_k)$ for compactness.

In a similar vein, it has been shown \cite{Kapec:2014opa,Campiglia:2014yka} that the subleading soft-graviton theorem is the Ward identity for the superrotations \cite{Barnich:2011ct} which generate an infinite-dimensional Virasoro subgroup of the extended BMS group\footnote{The sub-subleading soft-graviton theorem has also been recently recast as a symmetry of the $S$-matrix (see \cite{Campiglia:2016efb,Campiglia:2016jdj}).}.  In the language of 2D correlators, the current corresponding to these local conformal transformations is the stress tensor. We now turn to an explicit construction of this operator.

Our starting point is \eqref{zeromodeins2} which has a form reminiscent of a stress tensor Ward identity. To bring this into the usual form, we define
\begin{equation}
\begin{split}\label{Tzzdef}
T_{zz} \equiv \frac{i}{8\pi G} \int d^2 w \frac{1}{z - w } D_w^2 D^\bw N^\1_{\bw\bw} ~. 
\end{split}
\end{equation}
Then \eqref{zeromodeins2} implies 
\begin{equation}
\begin{split}\label{stresstensorins}
& \avg{ T_{zz}\CO_1 \cdots \CO_n   } =  \sum_{k=1}^n \left[   \frac{  h_k }{ ( z - z_k )^2 } + \frac{ \Gamma^{z_k}_{z_kz_k} }{z - z_k } h_k    +  \frac{1}{z - z_k }   \left( \p_{z_k} -   |s_k| \Omega_{z_k}   \right)  \right]  \avg{ \CO_1 \cdots \CO_n } ~ ,
\end{split}
\end{equation}
which is the precise form of the stress tensor correlator in a conformal field theory on a curved background. This can be brought to the more familiar form by dressing the operators with appropriate factors of the zweibein 
(see \cite{Eguchi:1986sb} for a more detailed discussion).

Define the charge 
\begin{equation}
\begin{split}\label{Tccharge}
T_\CC[Y] = \oint_\CC \frac{dz}{2\pi i} Y^z T_{zz} ~, 
\end{split}
\end{equation}
where $Y^z$ is a local CKV obeying $\p_\bz Y^z=0$ with no singularities inside the contour.
Insertions of \eqref{Tccharge} take the form
\begin{equation}
\begin{split}\label{Tcins}
& \avg{ T_\CC[Y]   \CO_1 \cdots \CO_n   } =  \sum_{k\in\CC} \left[    D_{z_k} Y^{z_k}  h_k   +    Y^{z_k} \left(\p_{z_k} -   |s_k| \Omega_{z_k}\right)  \right]  \avg{ \CO_1 \cdots \CO_n  } ~. 
\end{split}
\end{equation}
Thus, $T_\CC[Y]$ generates a local conformal transformation on all operators inside $\CC$\footnote{This operator is closely related to the soft part of the superrotation charge defined in \cite{Kapec:2014opa}. More precisely if $\CC$ is a contour that surrounds all $z_k$, then
\begin{equation} \notag
\begin{split}
Q^+_S = - \frac{i}{2} T_\CC[Y]  ~.
\end{split}
\end{equation}}.

Now, consider a contour $\CC$ that encircles all $z_k$ and a $Y^z$ that is globally defined on the sphere, i.e. $Y^z = a + b z + c z^2$. Since we are on a compact $S^2$, insertions of $T_\CC[Y]$ can be computed by either closing the contour towards $z=z_k$ or away from it. No poles are crossed when the contour is closed away from $z=z_k$ and these insertions must vanish. In other words, 
\begin{equation}
\begin{split}
\sum_{k=1}^n \left[    D_{z_k} Y^{z_k}  h_k   +    Y^{z_k} \left(\p_{z_k} -   |s_k| \Omega_{z_k}\right)   \right]  \avg{ \CO_1 \cdots \CO_n  } = 0 ~, \qquad Y^z = a +b z + c z^2 ~, 
\end{split}
\end{equation}
which is the statement of boost/angular momentum conservation \eqref{conslawexplicit}.

The stress tensor \eqref{Tzzdef} is non-local on $S^2$  in the news tensor zero mode $N^\1_{\bz\bz}$.  Nevertheless, we have proven that insertions of $T_{zz}$ are local on the $S^2$. 
 In contrast, the construction of the boundary stress tensor in AdS/CFT \cite{Brown:1992br,Balasubramanian:1999re} is local in the bulk fields when written in terms of subleading terms in the metric expansion. Leading and subleading terms in the metric expansion have a gauge-dependent and generally nonlocal relation on the $S^2$ enforced by the Einstein equation. We have tried but failed to find, by rewriting $N^\1_{\bz\bz}$ in terms of subleading metric components,  such a local expression in Bondi gauge\footnote{The $\mathcal{O}(r^0)$ term in $g_{zz}$ is an obvious suspect.}.
 However it is possible that such a manifestly local expression exists in some other gauge. 
 On the other hand, the nonlocality may indicate that the Virasoro action in 4D quantum gravity has a different character than that in conventional 2D CFT. 
 We leave this question unanswered for now.

Obviously an anti-holomorphic stress tensor $T_{\bz\bz}$ could be similarly constructed. However, a number of yet-unresolved issues arise for multiple soft-current insertions, even in the Maxwell case, as discussed in \cite{He:2014cra,He:2015zea}. The result of this paper is that insertions of a single $T_{zz}$ generate local conformal transformations when all other insertions are hard.

\section*{Acknowledgements}
We are grateful to T. Dumitrescu, J. Maldacena, S. Pasterski, B. Schwab, and A. Zhiboedov for useful discussions.  This work was supported in part by DOE grant DE-FG02-91ER40654.

\providecommand{\href}[2]{#2}\begingroup\raggedright\endgroup

%

\end{document}